\documentclass[aps,prl,twocolumn,nobibnotes,showpacs,superscriptaddress]{revtex4-1} 
\usepackage{graphicx} 
\usepackage{dcolumn} 
\usepackage{bm} 

\usepackage{amssymb} 
\usepackage{amsmath} 
\usepackage[alsoload=hep]{siunitx} 
\usepackage{color}
\usepackage{xcolor}
\usepackage{hyperref}

\graphicspath{{./}}

\usepackage{lineno}

\begin{document}

\title{Rosenbluth separation of  $ \boldsymbol {d\sigma_{L}/dt}$ and $\boldsymbol {d\sigma_{T}/dt}$ in $\boldsymbol \pi^{\bf 0}$ \\ deeply virtual electroproduction from the proton}

\author{I. Korover} \altaffiliation[On sabbatical leave from ]{Department of Physics, NRCN, P.O.B. 9001, Beer-Sheva 84190, Israel}
\affiliation{Laboratory for Nuclear Science, Massachusetts Institute of Technology, Cambridge, MA 02139.} 
\author{R.G. Milner} \affiliation{Laboratory for Nuclear Science, Massachusetts Institute of Technology, Cambridge, MA 02139.}

\date{\today}

\begin{abstract}
We report on a Rosenbluth separation using previously published data by the CLAS collaboration in Hall B, Jefferson Lab for exclusive $\pi^{0}$ deeply virtual electroproduction (DVEP) from the proton at a mean $Q^{2}$ of $\approx$ 2 (GeV/c)$^{2}$. The central question we address is the applicability of factorization in $\pi^0$ DVEP at these kinematics. The results of our Rosenbluth separation clearly demonstrate the dominance of the longitudinal contribution to the cross section. The extracted longitudinal and transverse contributions are in agreement with previous data from Hall A at Jefferson Lab, but over a much wider $-t$ range (0.12 - 1.8 (GeV/c)$^{2}$). The measured dominance of the longitudinal contribution at $Q^{2} \approx$ 2 (GeV/c)$^{2}$ is consistent with the expectation of the handbag factorization theorem. We find that $\sigma_L(t) \sim 1/(-t)$ for $-t >$ 0.5 (GeV/c)$^2$. Determination of both longitudinal and transverse contributions to the deeply virtual $\pi^{0}$ electroproduction cross section allows extraction of additional GPDs.
\end{abstract}

\maketitle

Hard lepton scattering from quarks is the elementary process by which we can understand the fundamental structure of matter in terms of the Standard Model.  The experimental discovery of deep inelastic scattering (DIS) by the MIT-SLAC Collaboration in 1967 and its theoretical interpretation by Bjorken and Feynman are the cornerstones for twenty-first century research to understand nucleon structure and properties.  \\

We now know that nucleons are many body interacting systems that are governed by the strong interaction. Exploring their structure is most effectively carried out experimentally using electron scattering and requires theoretical understanding of the measured observables in terms of the fundamental quark and gluon degrees of freedom of Quantum ChromoDynamics (QCD). The ability to explore the dynamics inside a nucleon relies on the factorization theorems, which provide a framework for the separation between hard and soft processes~\cite{Strikman:Factor,Factorization1,Factorization2}. These separations are proven to hold in the asymptotic limit when $Q^{2}\rightarrow\infty$, but there is some experimental evidence that this framework may hold as low as $Q^{2} = 1.5$ (GeV/c)$^{2}$ for Deeply Virtual Compton Scattering (DVCS)~\cite{Factorization}. In the Deeply Virtual Meson Production (DVMP) processes, the distinction between hard and soft contributions is somewhat ambiguous due to the presence of a meson in the final state. Factorization in DVMP processes should manifest itself by domination of the longitudinal cross section, $\sigma_{L}$ over the transverse cross section $\sigma_{T}$~\cite{Strikman:Factor}. The latter is expected to be suppressed by an additional factor $\frac{1}{Q^{2}}$. \\

\begin{figure}[!htpb]
\centering
\includegraphics[scale=0.2]{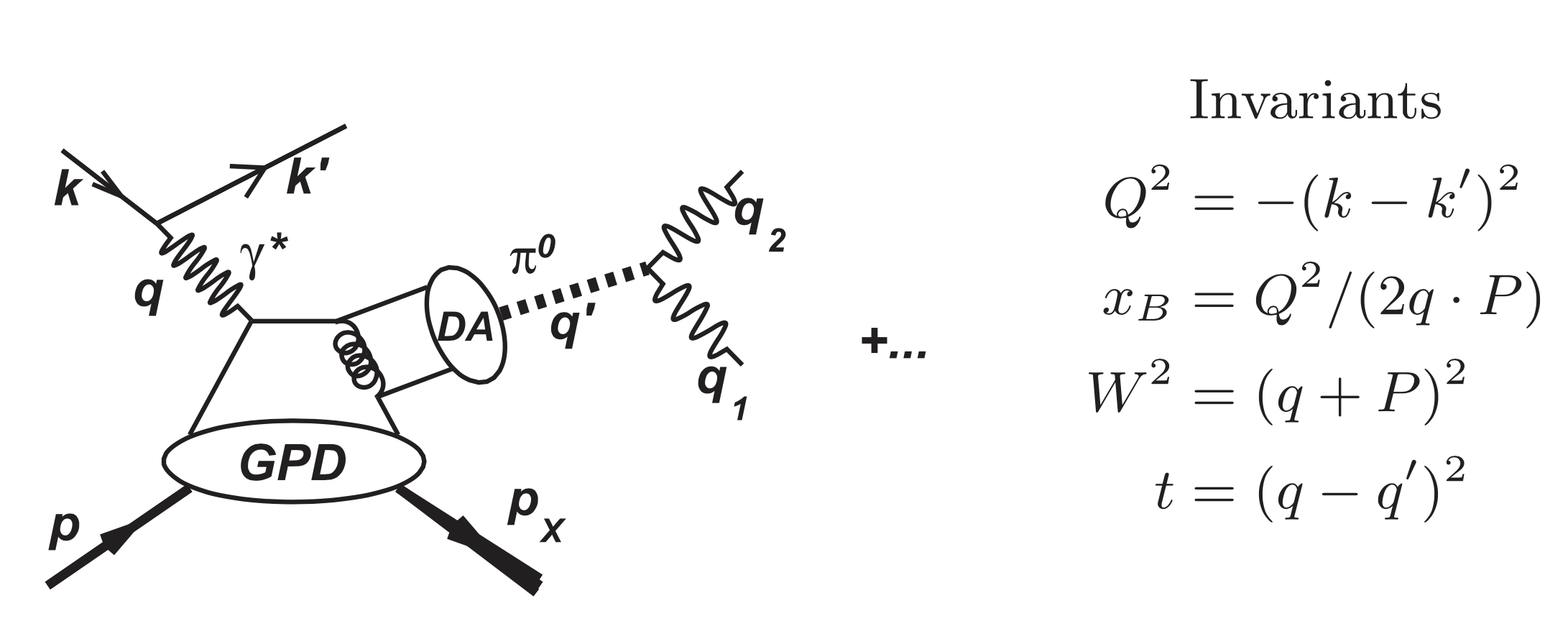}
\caption{Diagram of the exclusive $\pi^{0}$ electroproduction reaction, identified by the $\pi^{0}\rightarrow \gamma\gamma$ decay mode. The figure is adopted from~\cite{HallA}. The kinematic Lorentz invariants are defined on the right.}
\label{fig:handbag} 
\end{figure}

In general, according to the handbag approximation shown in Fig.~\ref{fig:handbag}, $\sigma_{L}$ is expected to scale as $\frac{1}{Q^{6}}$ while $\sigma_{T}$ scales as $\frac{1}{Q^{8}}$. Recent Rosenbluth separations of longitudinal and transverse cross sections were carried out with $\pi^{+}$ in the final state~\cite{piPlus,piPlus1}. In that case, the observed dominance of the longitudinal part at low values of $-t$ may in large part be due to the presence of the pion pole~\cite{piPlus}. To verify the kinematic region of the applicability of handbag factorization valuable information can be gathered from a process that is free from the pion pole contribution, e.g. deeply virtual $\pi^{0}$ exclusive electroproduction, and to extend the $-t$ range.
The leading twist approximation that is in good agreement with DVCS~\cite{dvcsFact} and vector meson production at high $Q^{2}$, underestimates the total $\pi^{0}$ electroproduction cross section in the collinear approximation by about one order of magnitude~\cite{underEstimate,BedlinskyPRC,BedlinskyPRL}. To explain this discrepancy for the neutral mesons, it was suggested that twist-3 quark helicity flip pion distribution amplitudes (DAs) coupled with the transversity GPDs of the proton would create a large cross section for transversely polarized virtual photons~\cite{twist3Mod}. These calculation were in agreement with the combined cross sections of the published CLAS data~\cite{BedlinskyPRC}, without violating the QCD factorization theorem.\\
In this work, we analyze CLAS data published in 2014~\cite{BedlinskyPRC} on exclusive electroproduction of $\pi^{0}$ mesons and extract separately the longitudinal and transverse cross sections for the available $-t$ range.
The DVMP cross section can be written in the following form~\cite{totalCross}:
\begin{equation}
\begin{split}
\frac{d^{4}\sigma}{dQ^{2}dx_{B} dt d\phi} = \frac{1}{2\pi} \frac{d^{2}\Gamma}{dx_{B}dQ^{2}}(Q^{2},x_{B},E) (\frac{d\sigma_{T}}{dt} + \varepsilon\frac{d\sigma_{L}}{dt}) + \\
	+ \sqrt{2\varepsilon(1+\varepsilon)}\frac{d\sigma_{TL}}{dt}\cos\phi + \varepsilon \frac{d\sigma_{TT}}{dt}\cos2\phi \ ,
\end{split}
\end{equation}
where the $(d^{2}\Gamma/dx_{B}dQ^{2})(Q^{2},x_{B},E)$ is the virtual photon flux, and $\varepsilon$ is the degree of longitudinal polarization of the virtual photon:

\begin{equation}
\varepsilon = \frac {1 - y - Q^{2}/4E^2}{1- y + y^2/2 + Q^2/(4E^{2})} \ ,
\end{equation} 
where  $y = (q\cdot P)/(k\cdot P)$.

Recent experiments conducted in Hall B by the CLAS Collaboration and  in Hall A at Jefferson Lab extracted the three combinations of structure functions, $(\frac{d\sigma_{T}}{dt} + \varepsilon\frac{d\sigma_{L}}{dt})$, $\frac{d\sigma_{TT}}{dt}$ and  $\frac{d\sigma_{LT}}{dt}$ by fitting the measured  $\phi_{\pi}$ distribution in each $(Q^{2},x_{B},t)$ bin to a functional form:
\begin{equation}
A + B\cdot \cos(\phi) + C\cdot \cos(2\phi) \ .
\end{equation}
However, in the published analysis the $\frac{d\sigma_{L}}{dt}$ and $\frac{\sigma_{T}}{dt}$ contributions were not separated, and only the combined $\frac{d\sigma_{T}}{dt} +\varepsilon \frac{d\sigma_{L}}{dt}$, $\frac{d\sigma_{LT}}{dt}$ and $\frac{d\sigma_{TT}}{dt}$ were reported.  We note that the Rosenbluth separation yielding $\frac{d\sigma_{L}}{dt}$ and $\frac{\sigma_{T}}{dt}$ contributions to deeply virtual exclusive $\pi^{0}$ electroproduction has been carried out in Hall A at Jefferson Lab. However, the small acceptance of the Hall A spectrometers limited the kinematic range accessible for DVMP. Moreover, the measured $-t$ range was limited to very low values. Although the extracted longitudinal and transverse cross sections were in good agreement with different models such as VGG, the results did not reproduce the interference terms.
On the other hand, the published CLAS data~\cite{BedlinskyPRC}, extend over a much larger kinematic range, but were not separated using the Rosenbluth method. Even if the CLAS data were taken at a single beam energy, it is possible to obtain an appreciable range in $\varepsilon$ by taking advantage of the large variation of scattering and polar angle in CLAS to reach near-equal values of $x_B$ and $Q^2$. Thus, the larger kinematic range of the CLAS data can be used to experimentally test the validity of the handbag factorization assumption. Moreover, by combining $\frac{d\sigma_{T}}{dt}$ and $\frac{d\sigma_{TT}}{dt}$ access to the GPDs is enhanced. 

\begin{figure}[!htpb]
\hspace*{-0.6cm}
\centering
\includegraphics[scale=0.18]{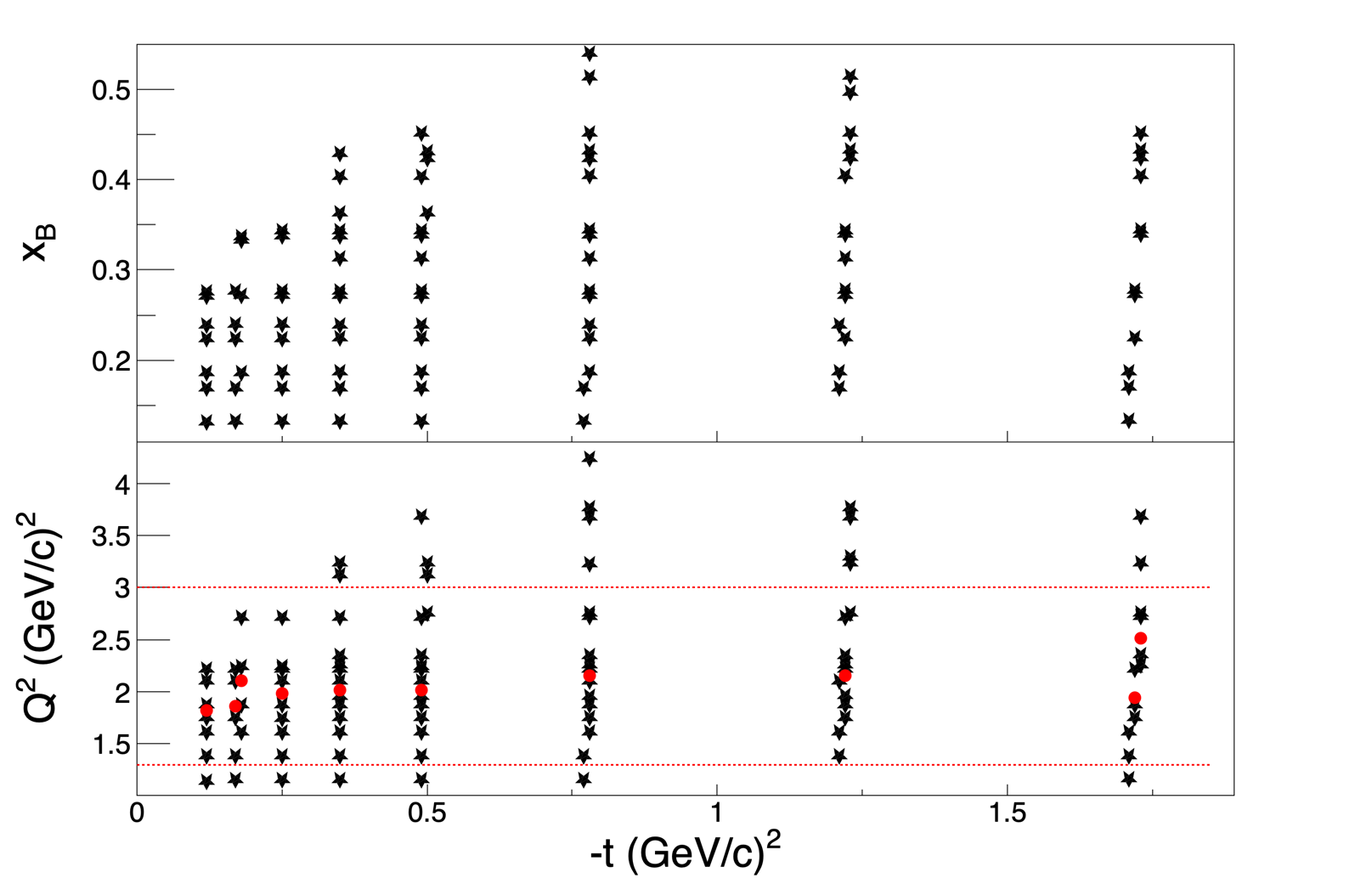}
\caption{$Q^{2}$ and $x_{B}$ coverage of the reported measurements as a function of $-t$. Stars indicate a specific $Q^{2}$ and $x_{B}$ value at each measurement, as reported in Table VIII in Ref.~\cite{BedlinskyPRC}. Horizontal red lines indicate a cut on minimal and maximum value of $Q^{2}$. The red circles indicate the mean value of $Q^{2}$, after applying $Q^{2}$ cuts.}
\label{fig:Kin_range} 
\end{figure}

The data used in the Rosenbluth separation reported here were taken using the CLAS detector, a 5.75 GeV electron beam and a 2.5 cm long liquid hydrogen target. The CLAS spectrometer was operated at an instantaneous luminosity of $2 \times 10^{34}$ cm$^{-2}$ s$^{-1}$ and $W>2$ GeV. All the data used in the Rosenbluth separation reported here were published and tabulated in~\cite{BedlinskyPRC}. The experimental coverage in $Q^{2}$ and $x_{B}$ is shown in Fig.~\ref{fig:Kin_range} as a function of $-t$ bins.

Due to the fact that the measurement was carried out at different values of $Q^{2}$ and $x_{B}$, we must verify that there is no additional hidden dependence on these kinematic quantities. 
 
\begin{figure}[t]
\hspace*{-0.6cm}
\centering
\includegraphics[scale=0.15]{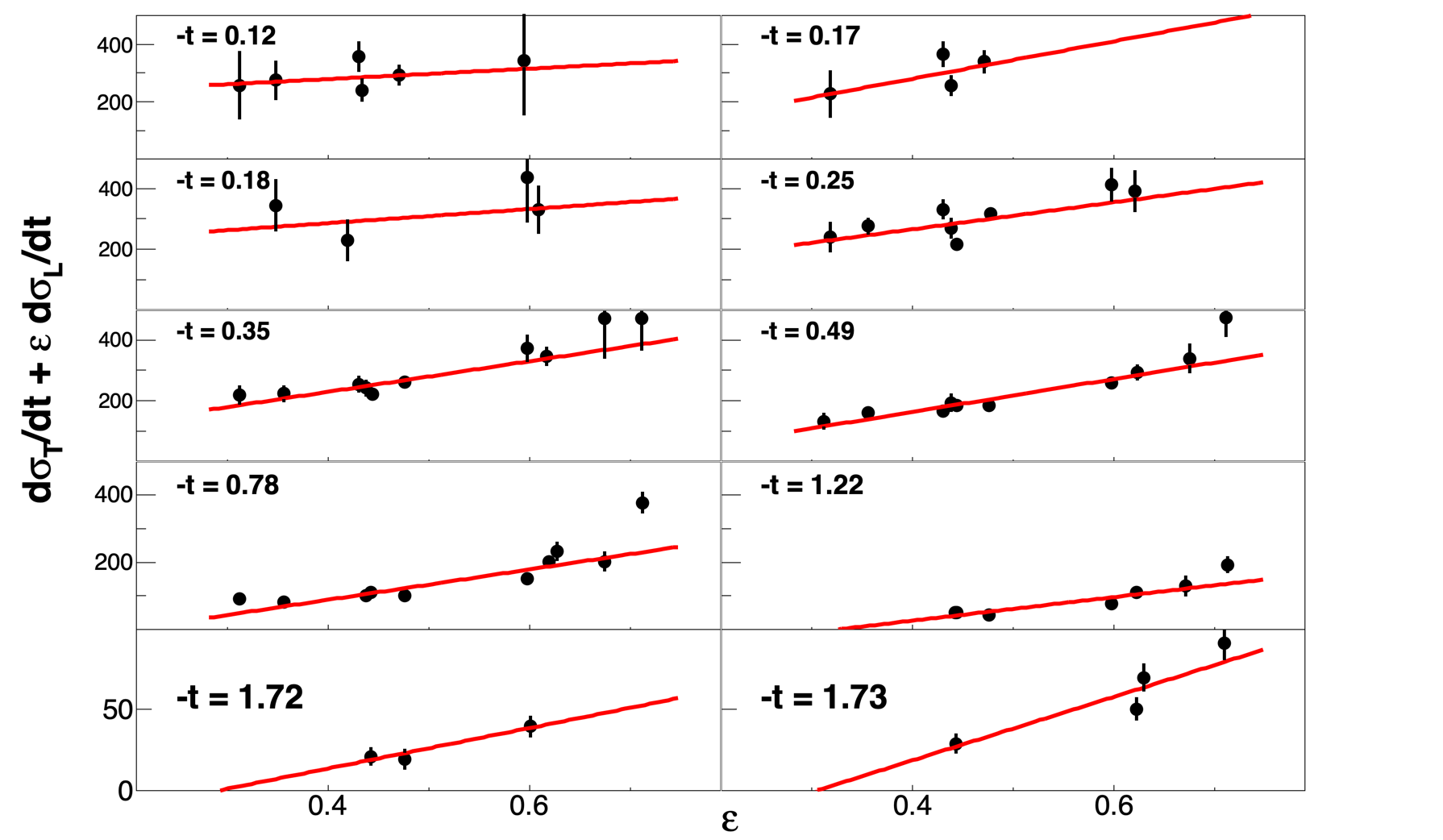}
\caption{Combined cross sections as a function of $\varepsilon$ for different $-t$ bins (0.12,0.17,0.18,0.25,0.35,0.49,0.78, 1.22,1.72,1.73 (GeV/c)$^{2}$) as labeled on the each pad. The red lines represent the linear fit to the data.}
\label{fig:RosSep}
\end{figure}

For each row of the tabulated experimental data, we calculated the $\varepsilon$ parameter and ploted the combined cross section $\frac{d\sigma_{T}}{dt} + \varepsilon\frac{ d\sigma_{L}}{dt}$ as a function of $\varepsilon$. This procedure was carried out for most available bins of $-t$. However, three $-t$ bins were not analyzed because they had less than three points in $\varepsilon$. The Rosenbluth separation fits to the combined cross section $\frac{d\sigma_{T}}{dt} + \varepsilon \frac{d\sigma_{L}}{dt}$ vs. $\varepsilon$ in the different $-t$ bins are shown in Fig.~\ref{fig:RosSep}.
The experimental data were fitted to linear functions without any additional constraints. The uncertainties at each data value were taken from Table VIII of~\cite{BedlinskyPRC}, and statistical and systematic uncertainties were combined linearly.

\begin{figure}[t]
\hspace*{-0.8cm}
\centering
\includegraphics[scale=0.25]{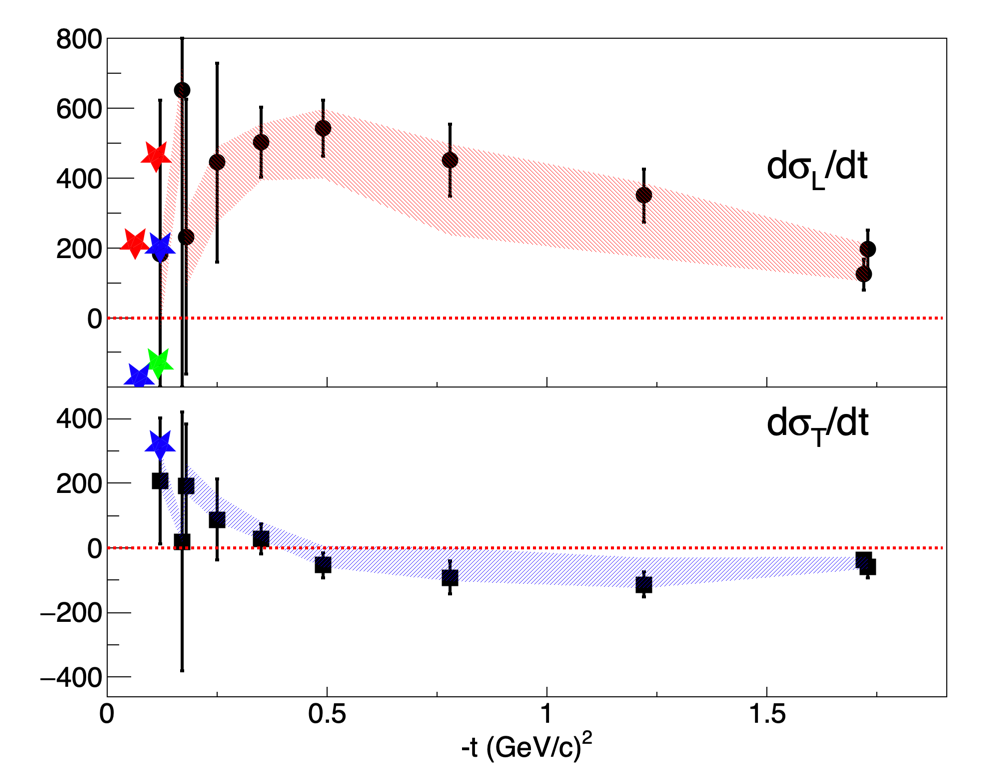}
\caption{Rosenbluth separated $\frac{d\sigma_{L}}{dt}$ and $\frac{d\sigma_{T}}{dt}$ as a function of $-t$. The black symbols represent this work. Circles (squares) are the extracted longitudinal $d\sigma_{L}/dt$ (transverse $d\sigma_{T}/dt$) for $Q^{2}_{min}>1.3$ GeV/c$^{2}$. The shaded red (blue) area represents the minimum and maximum range of the longitudinal (transverse) contributions to the cross section, following additional study of systematic uncertainties, as detailed below in the text. The star symbols represent Hall A data for different $Q^{2}$ bins: red (green, blue) stars for $Q^{2}=1.5$ (1.75, 2) GeV/c$^{2}$, and taken from~\cite{HallA}.}
\label{fig:final} 
\end{figure}

For the entire data set analyzed here, the average $Q^{2}$ is around 2 (GeV/c)$^{2}$ and the average $x_{B} \approx 0.3$.
Due to the limited $Q^{2}$ ($x_{B}$) range, the requirement of  minimal (maximal) value $Q^{2}_{min}$ ($Q^{2}_{max}$) did not change the mean significantly. However, even for the smallest available $Q^{2}$, the final results do not change significantly.\\
From the fits to the experimental data shown in Fig.~\ref{fig:RosSep}, we determine the longitudinal and transverse cross sections as a function of $-t$ for different requirements on $Q_{min}^{2}$. The results of the Rosenbluth separation, namely $\frac{d\sigma_{L}}{dt}$ and $\frac{d\sigma_{T}}{dt}$ vs. $-t$, are shown in separate panels of Fig.~\ref{fig:final} and are tabulated in Table~\ref{tab:data}. The uncertainties for the longitudinal and transverse contributions to the cross section are determined from an uncertainties of the slope and intersection point (with $\varepsilon$) given by fitting procedure. In addition for each point having $\frac{\chi^{2}}{ndf} > 1$, the uncertainties were multiplied by $\sqrt{\chi^{2}/ndf}$. 

The uncertainties in the separated longitudinal and transverse contributions are correlated. Thus, to extract the ratio of the longitudinal to transverse contributions with their correlated uncertainties, we have used the Monte Carlo method. For each $-t$ bin and at each $\varepsilon$, we generated the $\frac{d\sigma_{T}}{dt} + \varepsilon \frac{ d\sigma_{L}}{dt}$ values, by taking the experimental value as the mean and its uncertainty as the $1\sigma$ of a normal Gaussian distribution. For each realization, we fitted a linear function and extracted the longitudinal and transverse contributions. This Monte Carlo method gave very similar values of the separated cross sections to the standard fitting of Fig.~\ref{fig:RosSep} ($<1\%$ discrepancy).

However, the large angular acceptance of the CLAS detector may introduce additional systematic uncertainties that must be investigated. For example, the acceptance may introduce systematic point to point variations as well as  overall normalization uncertainties. 
We note that the original data were binned in $Q^{2}$ and $x_{B}$ and the mean values of the bins were used. 
We have investigated whether this procedure may introduce large variations in $\varepsilon$, depending on the position inside specific $Q^{2}$ and $x_{B}$ bins. 
We find that the extracted longitudinal and transverse contributions shown in Fig.~\ref{fig:final} are not very sensitive to the binning .\\
While the published CLAS data that were used here for the Rosenbluth separation~\cite{BedlinskyPRC} did include significant uncertainties, the negative values of $\frac{d\sigma_{T}}{dt}$ at large $-t$ that resulted from the separation, likely indicates that some residual uncertainties, beyond those included in the published analysis, exist and should be estimated.\\
This additional systematic uncertainty was estimated by carrying out the following studies:
\begin{itemize}
\item{A fit to the data was carried out where the normalization of each point of the published values of the combined cross section $\frac{d\sigma_{T}}{dt} + \varepsilon\frac{d\sigma_{L}}{dt}$ was varied by $\pm10\%$.}
\item{A fit to the data was carried out with the constraint that $\frac{d\sigma_{T}}{dt}$ not have a negative value.}
\item{A fit to the data was carried out with an $\varepsilon-$dependent correction to the points in Fig.~\ref{fig:RosSep} which changed the slope by up  by $10\%$.}
\item{A fit to the data was carried out with the constraint that $\varepsilon <0.6$ (except last two $-t$ settings, where $\varepsilon<0.64$ used, due to limited number of points). This cut removed forward going electrons. }
\end{itemize}

To summarize, the data points in Fig.~\ref{fig:final} are based on an unrestricted fit with propagation of the uncertainties as published.  The shaded areas indicate  the minimum and maximum values of the cross sections resulting from the above study of additional systematic effects. As can be seen from Fig.~\ref{fig:final}, even if one takes the full range of the possible variations (based on the existing data), the longitudinal contribution to the cross section is clearly dominant over the broad $-t$ range. We note that $\sigma_L(t) \sim 1/(-t)$ for $-t >$ 0.5 (GeV/c)$^2$.   

\begin{table}[!htpb]
\caption{Tabulated values of the longitudinal and transverse differential cross sections and their ratios for different $-t$ from the published CLAS data~\cite{BedlinskyPRC}.}
\begin{tabular}{c c c }
\hline
\hline
$-t$ (GeV/c)$^{2}$	&	 $d\sigma_{L}/dt$	 &	 $d\sigma_{T}/dt$	 \\
\hline
0.12 & 180 $\pm$ 443 & 207 $\pm$ 195	 \\
0.17 & 650 $\pm$ 909 & 20 $\pm$ 400	\\
0.18 & 232 $\pm$ 393 & 193 $\pm$ 190	 \\
0.25 & 444 $\pm$ 284 & 88 $\pm$ 126	 \\
0.35 & 503 $\pm$ 99 & 28  $\pm$  46	 \\
0.49 & 542 $\pm$ 79 & -53 $\pm$ 37 	 \\
0.78 & 451 $\pm$ 103 & -91 $\pm$ 50 	\\ 
1.22 & 350 $\pm$ 76 & -113 $\pm$ 38 	 \\
1.72 & 124 $\pm$ 44 & -36 $\pm$ 22 	 \\
1.73 & 195 $\pm$ 56 & -59 $\pm$ 33 	\\
\hline
\hline
\end{tabular}
\label{tab:data}
\end{table}

The $Q^{2}$ and $x_{B}$ kinematic range can impact the separation of longitudinal and transverse contributions. For example, the $\frac{d\sigma_{L}}{dt}$ cross section in the handbag approximation is expected to scale as $Q^{-6}$. With the available CLAS data~\cite{BedlinskyPRC}, it is not possible 
to form a conclusion regarding the $Q^{2}$ dependence of the longitudinal $\frac{d\sigma_{L}}{dt}$ contribution to the cross section.

For the charged-pion $\pi^+$ case, where separated data exist over a range in $Q^2$ up to $Q^2 \sim$ 4 (GeV/c)$^2$, the scaling laws were found to be reasonably consistent with the $Q^2$-dependence of the $\sigma_L$ data, but they failed to describe the $Q^2$-dependence of the $\sigma_T$ data~\cite{Horn:2007ug}. Rosenbluth-separated electroproduction data for K$^+$ are only available up to $Q^2 \sim$ 2.4 (GeV/c)$^2$, and here the $Q^2$-dependence for both $\sigma_L$ and $\sigma_T$ appear consistent with the QCD factorization prediction, albeit not conclusive due to the sparseness of the data~\cite{ Carmignotto:2018uqj}. At this modest $Q^2$ the $\sigma_T$ is still found to be dominant, with $\sigma_L$ and $\sigma_T$ near-equal at the highest $Q^2$.

We wish to emphasize that, by isolating the transverse contribution to the cross section $\frac{d\sigma_{T}} {dt}$ and combining it with the interference part $\frac{d\sigma_{TT}}{dt}$, $H_{T}$ and $\bar{E}_{T}$ GPDs can be isolated and studied as a function of $Q^{2}$, $-t$ and $x_{B}$.
 \begin{equation}
 \frac{d\sigma_{TT}}{dt} = \frac{4\pi\alpha}{k'}\frac{M^{2}_{\pi}}{Q^{8}} \frac{t'}{16m^{2}}|<\bar{E}_{T}>|^{2} \ .
 \end{equation}
  \begin{equation}
 \frac{d\sigma_{T}}{dt} = \frac{4\pi\alpha}{k'}\frac{M^{2}_{\pi}}{Q^{8}} [ (1-\xi)^{2} |<H_{T}>|^{2} - \frac{t'}{8m^{2}} |<\bar{E}_{T}>|^{2} ] \ .
 \end{equation}
In summary, from the analysis of the published CLAS data, we find that the longitudinal contribution to the cross section is dominant for $-t > 0.25$ (GeV)$^{2}$. This result, based on data taken at $Q^{2}\approx2$ (GeV/c)$^{2}$, is consistent with the prediction of factorization~\cite{Strikman:Factor} and the dominance of the handbag diagram of Fig.~\ref{fig:handbag}. Our results are also consistent with the previous Rosenbluth separation results from Hall A~\cite{HallA}, but we have been able to extend the experimental range to much larger $-t$ values. In the low $-t$ range, the most signiﬁcant contribution to the combined cross section is coming from the transverse contribution $\sigma_{T}$ as is also indicated in the Hall A data. However, when extending the range to higher $-t$ values, it is clear that the dominant part of the combined cross section arises mainly from the longitudinal contribution $\sigma_{L}$, while the transverse contribution is suppressed. This is consistent with the expectation based on the handbag diagram of Fig.~\ref{fig:handbag} and the factorization theorem of~\cite{Strikman:Factor}. We find that $\sigma_L(t) \sim 1/(-t)$ at high $-t$.   The dominance of $\sigma_L$ is robust to inclusion of significant, additional  systematic uncertainties beyond those assigned in the original published analysis. 
This work suggests that the factorization assumption associated with the handbag mechanism in neutral pion electroproduction can be applicable at $Q^{2}$ as low as 2 (GeV/c)$^{2}$. The limited $Q^{2}$ range of the CLAS data does not allow a more precise investigation of the $Q^{2}$ dependence. These conclusions strongly motivate precision Rosenbluth separations at high $-t$ of $\pi^0$ deeply virtual electroproduction from the proton. A dedicated E12-13-010 experiment to perform Rosenbluth separations at larger $Q^2$ but still relatively low $-t$ is approved to run in Hall C, taking advantage of the newly constructed Neutral Particle Spectrometer~\cite{Horn:2019beh}.
Further, the CLAS12 experiment E12-06-108, with the increased energy CEBAF will allow for measurement of deeply virtual neutral meson electroproduction over a somewhat larger kinematic range in $x_{B}$ and $Q^{2}$. In particular, it will be desirable to probe the $Q^{2}$ dependence of the longitudinal $\frac{d\sigma_{L}}{dt}$ contribution to the cross section. 
The future Electron Ion Collider (EIC) with its much larger $Q^{2}$ and substantially lower $x_{B}$ reach will probe nucleon structure in the region dominated by the sea quarks and gluons~\cite{EIC:white}.\\

We thank Rolf Ent, Or Hen, Tanja Horn and Mark Strikman for useful and insightful discussions.
This work was supported in part by the U.S. Department of Energy, Oﬃce of Nuclear Physics under contract DE-FG02-94ER40818.

\bibliography{reference}

\end{document}